\documentclass[pra,preprint, superscriptaddress, citeautoscript]{revtex4}

\usepackage{times}

\usepackage{amsfonts,amsmath,amssymb}
\usepackage[dvips]{graphicx}
\usepackage{dcolumn}
\usepackage{bm}
\newcounter{one}
\usepackage{braket}
\usepackage[usenames]{color}
\usepackage[tight]{subfigure}
\usepackage{float}
\usepackage[abs]{overpic}
\usepackage{pict2e}
\usepackage{color}

\begin{document}

\baselineskip24pt


\begin{center}
\bf
\large{Deterministic quantum teleportation of photonic quantum bits \\ by a hybrid technique}
\end{center}
\begin{center}
\normalsize{
Shuntaro Takeda$^{1}$, Takahiro Mizuta$^{1}$, Maria Fuwa$^{1}$, Peter van Loock$^{2}$, \& Akira Furusawa$^{1}$
}
\end{center}
\begin{center}
\it
\normalsize{
$^{1}$Department of Applied Physics, School of Engineering, The University of Tokyo,\\ 7-3-1 Hongo, Bunkyo-ku, Tokyo 113-8656, Japan\\
$^{2}$Institute of Physics, Johannes-Gutenberg Universit\"at Mainz, Staudingerweg 7,\\ 55128 Mainz, Germany
}
\end{center}

\vspace{\baselineskip}


{\bf
Quantum teleportation~\cite{93Bennett} allows for the transfer of
arbitrary, in principle, unknown quantum states from a sender to a
spatially distant receiver, who share an entangled state and can communicate classically.
It is the essence of many sophisticated protocols for quantum communication
and computation~\cite{98Briegel,99Gottesman,01Knill,01Raussendorf}.
In order to realize flying qubits in these schemes, photons are an optimal choice,
however, teleporting a photonic qubit~\cite{97Bouwmeester,98Boschi,01Kim,03Marcikic,03Pan,12Ma}
has been limited due to experimental inefficiencies and restrictions.
Major disadvantages have been the fundamentally probabilistic nature of linear-optics
Bell measurements~\cite{Luetkenhaus} as well as the need for either destroying the teleported qubit
or attenuating the input qubit when the detectors do not resolve photon numbers~\cite{12Pan}.
Here we experimentally realize fully deterministic, unconditional quantum teleportation of
photonic qubits. The key element is to make use of a ``hybrid'' technique:
continuous-variable (CV) teleportation~\cite{94Vaidman,SamKimble,98Furusawa} of a discrete-variable, photonic qubit.
By optimally tuning the receiver's feedforward gain, the CV teleporter acts
as a pure loss channel~\cite{01Hofmann,Polkinghorne}, while the input dual-rail encoded qubit, based on a single photon, represents
a quantum error detection code against amplitude damping~\cite{NielsenChuang}
and hence remains completely intact for most teleportation events.
This allows for a faithful qubit transfer even with imperfect CV entangled states:
the overall transfer fidelities
range from 0.79 to 0.82 for four distinct qubits, all of them
exceeding the classical limit of teleportation.
Furthermore, even for a relatively low level of the entanglement,
qubits are teleported much more efficiently than in previous experiments, albeit post-selectively
(taking into account only the qubit subspaces),
with a fidelity comparable to the previously reported values.
}

\vspace{\baselineskip}


Since its original proposal by Bennett \textit{et al.}~\cite{93Bennett},
the concept of quantum teleportation has attracted a lot of attention and has even become one of the central elements for advanced and practical realizations of quantum information protocols.
It is essential for long-distance quantum communication by means of quantum repeaters~\cite{98Briegel}
and it has also been shown to be a useful tool for realizing universal quantum logic gates
in a measurement-based fashion~\cite{99Gottesman}.
Many proposals and models for quantum computation rely upon quantum teleportation,
such as the efficient linear-optics quantum computing scheme by Knill, Laflamme, and Milburn~\cite{01Knill}
and the so-called one-way quantum computer using cluster states~\cite{01Raussendorf}.

Although much progress has been made in demonstrating quantum teleportation of photonic qubits~\cite{97Bouwmeester,98Boschi,01Kim,03Marcikic,03Pan,12Ma}, most of these schemes
shared one fundamental restriction: an unambiguous two-qubit Bell-state measurement (BSM), as needed
for teleporting a qubit using two-qubit entanglement, is always probabilistic
when linear optics is employed, even if photon-number-resolving detectors (PNRDs) had been available~\cite{Luetkenhaus,12Pan}.
There are two experiments avoiding this constraint, however, in these, either a qubit can no longer be teleported when it is coming independently from the outside~\cite{98Boschi}
or an extra nonlinear element leads to extremely low measurement efficiencies of the order of $10^{-10}$~\cite{01Kim}.
A further experimental limitation, rendering these schemes fairly inefficient,
is the probabilistic nature of the entangled resource states~\cite{12Pan}.
Efficient, near-deterministic quantum teleportation, however, is of great benefit in quantum communication
in order to save quantum memories in a quantum repeater; and it is a necessity in teleportation-based
quantum computation. An additional drawback of the previous experiments, due to the lack of PNRDs,
was the need for either
destroying the teleported qubit~\cite{98Braunstein} or attenuating the input qubit~\cite{03Pan},
thus further decreasing the success rate of teleportation.

We overcome all the above limitations through a totally distinct approach:
continuous-variable (CV) quantum teleportation of a photonic qubit.
The strength of CV teleportation lies in the on-demand availability of the quadrature-entangled states
and the completeness of a BSM in the quadrature bases using linear optics and homodyne detections~\cite{SamKimble}.
So far, these tools have been employed to unconditionally teleport CV quantum states such as coherent states~\cite{98Furusawa,07Yonezawa}.
However, their application to qubits~\cite{01Ide,Polkinghorne}
has long been out of reach,
since typical pulsed-laser-based qubits (like those in Refs.~~\cite{97Bouwmeester,98Boschi,01Kim,03Marcikic,03Pan,12Ma})
have a broad frequency bandwidth, incompatible with
the original continuous-wave-based CV teleporter that only works on narrow sidebands~\cite{98Furusawa,07Yonezawa}.
We overcome this incompatibility by utilizing a very recent, advanced technology:
a broadband CV teleporter~\cite{11Lee} and a narrow-band time-bin qubit compatible with that teleporter~\cite{12Takeda2}.
Importantly, this qubit uses two temporal modes to represent a so-called dual-rail encoded qubit~\cite{12Pan},
\begin{align}
\ket{\psi}=\alpha\ket{0,1}+\beta\ket{1,0},
\label{eq:qubit}
\end{align}
where $\ket{0,1}$ and $\ket{1,0}$ represent a photon in either of the temporal modes (expressed in the two-mode photon-number basis, $|\alpha|^2+|\beta|^2=1$).
Therefore, teleportation of both rails of the qubit is accomplished by means of a single CV teleporter
acting subsequently on the temporal modes of the time-bin qubits (Fig.~1).

Remarkably, the main weakness of CV teleportation, namely the intrinsic imperfection of the finitely squeezed, entangled states, can be circumvented to a great extent
in the present ``hybrid'' setting when the input to the CV teleporter is a dual-rail qubit.
The entangled state of the CV teleporter is a two-mode squeezed,
quadrature-entangled state,
$\sqrt{1-g_\text{opt}^2}\sum_{n=0}^{\infty}g_\text{opt}^n \ket{n,n}$;
here written in the number basis,
with $g_\text{opt}\equiv\tanh r$ and a squeezing parameter $r$.
Since infinite squeezing ($r\to\infty$) requires infinite energy,
maximally entangled states are physically unattainable, and thus,
the teleportation fidelity is generally limited by the squeezing level $r$.
Following the standard CV quantum teleportation protocol with
unit gain for the receiver's feedforward displacement~\cite{SamKimble}
yields a largely distorted output qubit with additional thermal photons.
In contrast, non-unit gain conditions are useful in some cases~\cite{03Bowen,04Jia}.
Especially, a single-mode CV teleporter with gain $g_\text{opt}$ creates no additional photons,
since it is equivalent to a pure attenuation channel with an intensity fraction of $(1-g_\text{opt}^2)$ getting lost into the environment~\cite{01Hofmann,Polkinghorne}.
Moreover, the dual-rail qubit basis spans a quantum error detection code
against such amplitude damping, where either a photon-loss error occurs, erasing the qubit, or otherwise
a symmetric damping leaves the input qubit state completely intact \cite{NielsenChuang}.
These two facts together mean that the dual-rail CV teleporter at optimal gain $g_\text{opt}$
transforms the initial qubit state as
\begin{align}
\ket{\psi}\!\bra{\psi}\longrightarrow g_\text{opt}^2\ket{\psi}\!\bra{\psi}+(1-g_\text{opt}^2)\ket{0,0}\!\bra{0,0}.
\label{eq:optimal_gain_teleportation}
\end{align}
Most importantly, no additional photons are created and the quantum information encoded into $\ket{\psi}$ remains undisturbed regardless of the squeezing level. The only effect of the teleporter
is the extra two-mode vacuum term, whose fraction would become arbitrarily small for sufficiently
large squeezing, $g_\text{opt}\to 1$.
This technique allows us to teleport arbitrary qubit states more faithfully by suppressing additional photons,
thereby realizing unconditional teleportation with a moderate level of squeezing.
Equation~(\ref{eq:optimal_gain_teleportation}) also shows that a fidelity of unity is obtainable
for any nonzero squeezing level, $g_\text{opt}> 0$, provided the signal qubit subspace is
post-selected, i.e., the non-occurrence of a photon-loss error is detected with a probability
approaching zero for $g_\text{opt}\to 0$.
We note that the remaining vacuum contribution
could be made arbitrarily small also without post-selection of the final states, by instead
immediately discarding all those quadrature results of the BSM which are too far from the phase-space origin~\cite{01Ide,10Mista}.

In order to demonstrate successful qubit quantum teleportation,
we prepare four distinct qubit states:
$\ket{0,1}$, $\ket{1,0}$,
$\ket{\psi_1}\equiv(\ket{0,1}-i\ket{1,0})/\sqrt2$, and
$\ket{\psi_2}\equiv(2\ket{0,1}-\ket{1,0})/\sqrt5$.
This set, including even and uneven superpositions of $\ket{0,1}$ and $\ket{1,0}$
with both real and imaginary phases, represents a fair sample of
qubit states on the Bloch sphere. In theory, our teleporter acts on any qubit state
in the same way (see supplementary discussion).

The experimental density matrix of the input state $\ket{\psi_1}$ is shown in Fig.~2(a).
This input state is not a pure qubit state, but rather a mixed state with
a $25\pm1\%$ vacuum, a $69\pm1\%$ qubit, and a $6\pm1\%$ multi-photon contribution.
Since the CV teleporter transfers input states of arbitrary dimension,
all these components are teleported and constitute the final, mixed output state.
Note that for our first analysis, we do not discard any of these contributions
from the input or the output states, thus ensuring that none of the quantum states
that enter or leave our teleporter are pre-selected or post-selected, respectively.

First we present the output state of unit-gain teleportation with $r=1.01\pm0.03$ in Fig.~2(b).
All the matrix elements obtained are in good agreement with theory:
the qubit term fraction drops, while the contribution of the multi-photon terms grows due to the finite squeezing.
The off-diagonal elements of the qubit ($\ket{0,1}\!\bra{1,0}$, $\ket{1,0}\!\bra{0,1}$) clearly retain
the original phase information of the input superposition between $\ket{0,1}$ and $\ket{1,0}$,
demonstrating that the non-classical feature of the qubits is preserved during the teleportation process.
These off-diagonal elements, however, decay a little more rapidly compared to the diagonal elements ($\ket{0,1}\!\bra{0,1}$, $\ket{1,0}\!\bra{1,0}$), illustrating
that the quantum superposition of the qubit is the most fragile feature in an experimental situation.

Next we turned down the gain $g$ and observed the new output state.
Figure~2(c) shows the output state at $g=0.79$ (close to $g_\text{opt}=0.77$).
Compared to Fig.~2(b), this time it can be seen that
the qubit components are almost undisturbed, while the vacuum grows
and the occurrence of extra multi-photon components is suppressed.
Thus, here the input-output relation is similar to the pure-attenuation model with a loss fraction of $1-g_\text{opt}^2=0.41$.
The bar graph in Fig.~3 shows the $g$ dependence of the
qubit/multi-photon components in the output state,
clearly demonstrating that gain-tuning reduces the creation of additional photons in CV teleportation.

The performance of teleportation can be assessed by means of the fidelity.
In our deterministic scheme, we must take into account the vacuum and multi-photon contributions,
unlike previous non-deterministic teleportation experiments using post-selection.
The overall fidelity between the input state $\hat{\rho}_\text{in}$ and the output state $\hat{\rho}_\text{out}$
is~\cite{94Jozsa}
\begin{align}
F_\text{state}=\left[\text{Tr}\left(\sqrt{\sqrt{\hat{\rho}_\text{in}}
\hat{\rho}_\text{out}\sqrt{\hat{\rho}_\text{in}}}\right)\right]^2.
\end{align}
When $\hat{\rho}_\text{in}$ has a qubit fraction $\ket{\psi}\!\bra{\psi}$ of $\eta$,
the classical bound for $F_\text{state}$ corresponds to $F_\text{thr}\equiv1-\eta/3$,
which is the best fidelity achievable without entanglement
(see supplementary discussion).
Therefore, $F_\text{state}>F_\text{thr}$ is a rigorous success criterion of
unconditional quantum teleportation.
Alternatively, we may also assess our teleporter
by calculating the fidelity after post-selecting the qubit components alone:
$F_\text{qubit}=\braket{\psi|\hat{\rho}_\text{out}^\text{qubit}|\psi}$,
where $\ket{\psi}$ is the ideal qubit state and
$\hat{\rho}_\text{out}^\text{qubit}$ is obtained by
extracting and renormalizing the qubit subspace spanned by $\{\ket{0,1},\ket{1,0}\}$ from the output density matrix.
Note that $F_\text{qubit}>2/3$ is known to be the success criterion of post-selective teleportation
with a pure input qubit and a mixed output qubit~\cite{95Massar}.

As shown in Fig.~3,
the $g$ dependence of these two fidelities is in good agreement with the theoretical predictions.
The maximal fidelities are obtained at $g=0.79$. Most importantly,
here, we do not only satisfy the usual qubit-subspace teleportation criterion, $F_\text{qubit}=0.875\pm0.015>2/3$,
but also the fully non-post-selected, Fock-space criterion, $F_\text{state}=0.817\pm0.012>F_\text{thr}=0.769\pm0.004$, thus demonstrating deterministic, unconditional quantum teleportation of a photonic qubit.
Besides the input qubit $\ket{\psi_1}$,
the Fock-space criterion is also fulfilled for the other three qubit states $\ket{0,1}$, $\ket{1,0}$, and $\ket{\psi_2}$
with the same experimental $r$ and $g$ values,
where $F_\text{state}=0.800\pm0.006$, $0.789\pm0.006$, and $0.796\pm0.011$ are observed, respectively
(theoretically, $F_\text{state}$ and $F_\text{qubit}$ are independent of the qubit; see supplementary discussion and data).
Note that, although the pure-attenuation model predicts $F_\text{qubit}=1$ and a complete suppression of multi-photon terms at gain $g_\text{opt}$,
our results slightly deviate from that situation due to experimental imperfections, such as extra loss and phase fluctuations of the squeezing.

Finally, Fig.~2(d) shows the output state for the lower squeezing level $r=0.71\pm0.02$ and $g=0.63$ ($g_\text{opt}=0.61$).
Here, although the vacuum component becomes more dominant,
the qubit components still retain almost the same form as in Fig.~2(c).
Under these circumstances,
the success of teleportation is only post-selective ($F_\text{qubit}=0.879\pm0.015>2/3$, $F_\text{state}<F_\text{thr}$) due to the insufficient squeezing resource.
However, the overall success probability for transferring the qubits ($43\pm3\%$, the ratio of the input and output qubit components)
is still much higher than that of previous experiments (far below $1\%$), even for this relatively low squeezing level.
This clearly shows the big advantage of our hybrid approach over the standard approaches.

In conclusion, we experimentally realized unconditional quantum
teleportation of four distinct
photonic qubit states, beating the fidelity
limits of classical teleportation in a deterministic fashion. In our scheme,
once the input qubit states are prepared, there is no need for pre-processing or
post-selecting them, and the teleported states freely emerge at the output of our teleporter.


\section*{METHODS}

Our experimental setup is shown in Fig.~1.
The time-bin qubit is conditionally created at a rate of $\sim 5000$~s$^{-1}$ from a continuous-wave laser~\cite{12Takeda2}
(wavelength: 860 nm), by extending the technique of Ref.~\cite{06Zavatta}.
Each time-bin has a frequency bandwidth of 6.2 MHz around the laser frequency.
Our CV teleporter~\cite{11Lee} operates continuously with a bandwidth of 12 MHz around the laser frequency,
which is sufficiently wide to cover the qubit bandwidth -- ultimately enabling us to teleport qubits
in a deterministic fashion.
In our CV teleporter, two single-mode squeezed states (each with an ideal, pure squeezing parameter $r$)
from two optical parametric oscillators (OPO) are suitably mixed at a 50:50 beam splitter (BS) to generate the quadrature-entangled beams.
This entanglement source is permanently available with no need for any probabilistic
heralding mechanism.
At the sending station of the teleporter, the input qubit is first combined with one half of the entangled beams at a 50:50 BS.
A complete CV BSM is then performed by measuring the two output modes of the BS through
two homodyne detections of two orthogonal quadratures.
These homodyne signals are classically communicated to the receiving station, where they are
multiplied with a gain factor (gain: $g$) and fedforward by means of
a displacement operation on the other half of the entangled beams.
Time synchronization of this final displacement is achieved by introducing an optical delay
to the corresponding entangled beam.
Finally, the output state is characterized via
single- or dual-homodyne measurement~\cite{12Takeda2}.
For every state, 100000 sets of quadrature values are recorded and the corresponding
two-mode density matrix is reconstructed by means of maximum-likelihood technique
without compensating finite measurement efficiencies.


\hspace{\baselineskip}


\noindent
{\bf Acknowledgements} \quad
This work was partly supported by the SCOPE program of the MIC of Japan, PDIS, GIA, G-COE, APSA, and FIRST commissioned by t
he MEXT of Japan, and ASCR-JSPS.
S. T. and M. F. acknowledge financial support from ALPS.
We thank Ladislav Mi\v{s}ta Jr., Hidehiro Yonezawa, and Jeff Kimble for useful comments.

\noindent
{\bf Author Contributions} \quad
A.F. planned and supervised the project.
P.v.L. and S.T. theoretically defined scientific goals.
S.T. and T.M. designed and performed the experiment, and acquired data.
S.T. and M.F. developed electric devices.
S.T., T.M. and M.F. analyzed data.
S.T. and P.v.L. wrote the manuscript with assistance from all other co-authors.

\noindent
{\bf Author Information} \quad
Reprints and permissions information is available at www.nature.com/reprints.
The authors declare no competing financial interests.
Correspondence and requests for materials should be addressed to A.F.
(akiraf@ap.t.u-tokyo.ac.jp).


\clearpage

\begin{figure}[!ht]
\begin{center}
\includegraphics[width=1.0\linewidth]{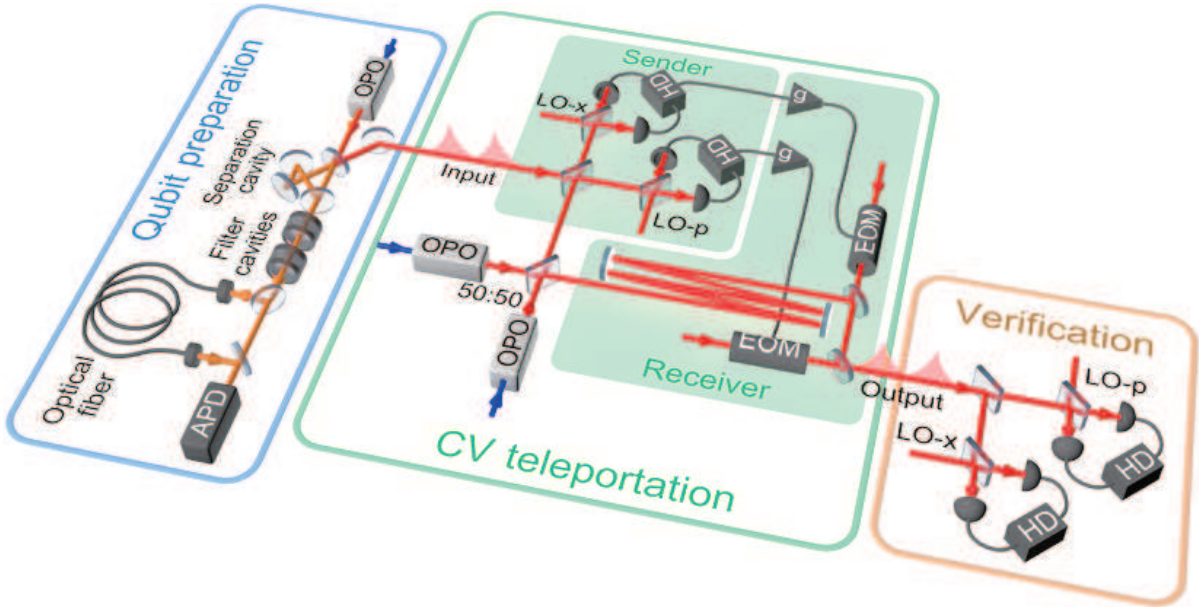}
\end{center}
\caption{{\bf Experimental setup.} A time-bin qubit is heralded by detecting one half of an entangled photon pair
produced by an OPO.
The CV teleporter ($g$: feedforward gain) always transfers this qubit, albeit with non-unit fidelity.
The teleported qubit is finally characterized by
single- or dual-homodyne measurement to verify the success
of teleportation. See Methods for details.
APD; avalanche photo-diode, EOM; electro-optic modulator, HD; homodyne detector, and LO; local oscillator.}
\label{fig:schematic}
\end{figure}

\begin{figure}[!ht]
\centering
\includegraphics[clip,scale=1.0]{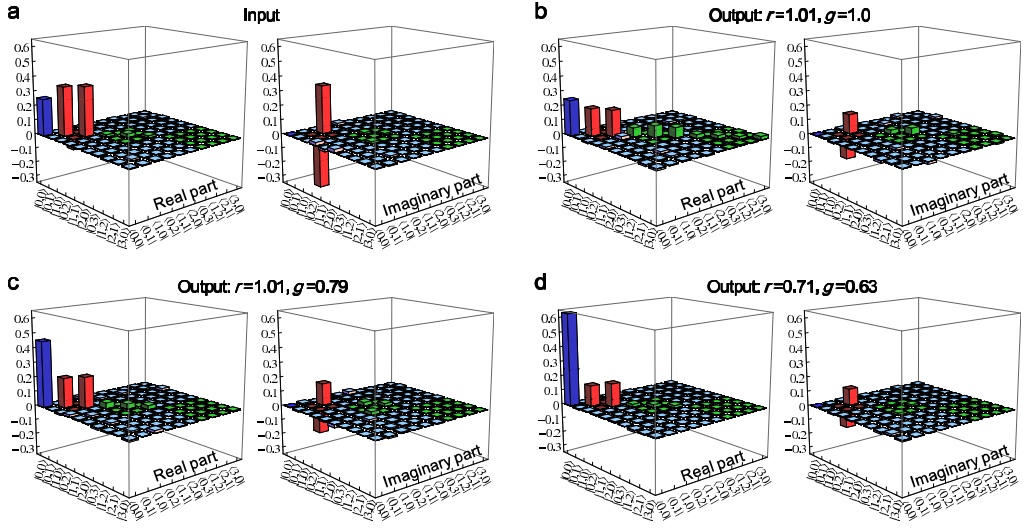}
\caption{{\bf Experimental density matrices.}
By means of homodyne tomography, two-mode density matrices are reconstructed both for the input and the output states in photon-number bases~\cite{12Takeda2}:
$\hat{\rho}=\sum_{k,l,m,n=0}^\infty\rho_{klmn}\ket{k,l}\!\bra{m,n}$.
The bars show the real or imaginary parts of each matrix element $\rho_{klmn}$.
Blue, red, and green bars correspond to the vacuum, qubit, and multi-photon components, respectively.
{\bf a.} Input state $\ket{\psi_1}$.
{\bf b-d.} Output states for different $r$ and $g$.}
\label{fig:density_matrix3}
\end{figure}

\begin{figure}[!ht]
\centering
\includegraphics[clip,scale=1.0]{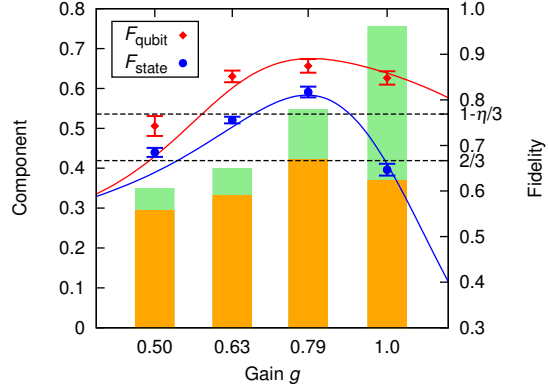}
\caption{{\bf Experimental results of teleportation including gain tuning.}
The horizontal axis $g$ uses a logarithmic scale.
Orange and green bars represent, respectively, qubit and multi-photon components
of the teleported states (the left vertical axis).
Red diamonds and blue circles with error bars ($\pm1$ standard deviation) correspond to $F_\text{qubit}$ and $F_\text{state}$, respectively (the right vertical axis).
Theoretical fidelity curves (see supplementary information) are also plotted with the same colors.
All observed $F_\text{qubit}$ significantly exceed the classical limit of 2/3.
At $g=0.79$, also $F_\text{state}>1-\eta/3$ is satisfied and thus unconditional teleportation is demonstrated.}
\label{fig:g_dependence2}
\end{figure}

\noindent

\clearpage



\begin{center}
\bf
\large{Supplementary Information}
\end{center}

\section{Supplementary Discussion}

\subsection{Success criterion}

We shall derive a success criterion for quantum teleportation
applicable to the imperfect, heralded dual-rail qubits
used in our experiment.

The experimental input state to our CV teleporter
cannot be represented by a pure, dual-rail qubit state.
Instead, it should be more generally
expressed by an initial two-mode density operator of the form
\begin{align}
\hat{\rho}_{\text{in}}=\eta\ket{\psi}\!\bra{\psi}+(1-\eta)\,\hat\rho^\perp\,,
\end{align}
where $\ket{\psi}$ is a pure qubit state encoded into the dual-rail
basis $\{\ket{01},\ket{10}\}$ and the operator $\hat\rho^\perp$ contains
all those two-mode Fock-space terms orthogonal to $\{\ket{01},\ket{10}\}$,
i.e., multi-photon terms and, in particular, the two-mode vacuum term
$\ket{00}\!\bra{00}$ (compared to the main text, for notational convenience,
here we drop the commas in the two-mode state vectors).
This mixed state is given to the sender of the teleporter (Alice) who
is supposed to transfer it as reliably as possible to the receiver (Bob).
For such input states,
we would like to know when it is justified to claim that
the entanglement-assisted quantum teleporter operates in a regime
inaccessible to a classical teleporter.
In classical teleportation, Alice and Bob do not share an entangled state.

In order to derive the optimal fidelity in a classical teleporter
(that makes no use of entanglement), it is important to
notice that, while $\ket{\psi}$ is an arbitrary qubit state unknown
to Alice and Bob, the state
$\hat\rho^\perp$, whatever complicated and high-dimensional it may be,
is, in principle, perfectly known.
Thus, when looking at the two orthogonal subspaces (qubit and non-qubit) of
$\hat{\rho}_{\text{in}}$ separately, we obtain
the following bounds for classical teleportation:
\begin{itemize}
\item $F=1$ for the case of teleporting $\hat\rho^\perp$\,,
\item $F=2/3$ for the case of teleporting $\ket{\psi}$~\cite{s95Massar}.
\end{itemize}
Intuitively, we may then expect an overall classical bound such as
\begin{align}
F=\eta\cdot\frac23+(1-\eta)\cdot1=1-\frac{\eta}3\,,
\label{eq:threshold_1}
\end{align}
corresponding to an average fidelity for (optimally) classically teleporting
either the unknown pure qubit state or the known mixed, orthogonal complement of it.
More formally, however, the classical fidelity limit should be obtained by
comparing the total mixed input state $\hat{\rho}_\text{in}$
with the total mixed output state $\hat{\rho}_\text{out}$ through the
mixed-state fidelity~\cite{s94Jozsa}
\begin{align}
F_\text{state}=\left[\text{Tr}\left(\sqrt{\sqrt{\hat{\rho}_\text{in}}
\hat{\rho}_\text{out}\sqrt{\hat{\rho}_\text{in}}}\right)\right]^2.
\end{align}

In order to derive the desired classical bound, let us
make two assumptions that greatly simplify the analysis
without loss of generality.
First, we can set $\ket{\psi}=\ket{01}$, bearing in mind that
this state is still unknown to Alice.
Further, we assume $\hat\rho^\perp\equiv \ket{00}\!\bra{00}$,
noting that the derivation below would follow through unaltered
even if additional non-qubit terms were present in $\hat\rho^\perp$,
as long as $\hat\rho^\perp$ remains known and orthogonal to
the qubit subspace. Hence, without loss of generality,
the input density matrix written in the two-mode photon-number
basis $\{\ket{00},\ket{01},\ket{10}\}$ is
\begin{align}
\rho_{\text{in}}=
\left[\begin{array}{c|cc}
1-\eta & 0 & 0 \\ \hline
0 & \eta & 0 \\
0 & 0 & 0 \\ \end{array}\right]\,.
\label{eq:rho_in}
\end{align}
One possible strategy for Alice is now to
use quantum non-demolition(QND)-type photon measurements,
in order to determine whether the input state is in one of the two subspaces
$\{\ket{00}\}$ (with probability $1-\eta$) or
$\{\ket{01},\ket{10}\}$ (with probability $\eta$). Since this strategy
leaves the entire quantum information contained in
$\hat{\rho}_\text{in}$ completely intact
(i.e., even an arbitrary qubit state $\ket{\psi}$ will be preserved
in the subspace $\{\ket{01},\ket{10}\}$), it constitutes
a possible first step for an optimal classical scheme.
Once the corresponding subspace is determined, optimal further steps
can be easily found. All these steps, as usual in classical teleportation,
rely upon local quantum state estimation by Alice, classical
communication of the estimate from Alice to Bob,
and local state preparation by Bob according to Alice's classical message.
In the following, we shall refer to the last two steps simply
as Alice sending the corresponding estimated (or guessed) quantum state to Bob
(bearing in mind that typically,
a perfect quantum channel for an actual direct transmission of quantum states
will not be available, as otherwise Alice could instead send the
unknown input state directly to Bob with no need for any state estimation).

First, assume Alice obtains $\ket{00}\!\bra{00}$ with probability $1-\eta$.
In this case,
Alice will send Bob either a vacuum state $\ket{00}\!\bra{00}$ with probability $1-x$
or a randomly chosen, guessed qubit $\ket{\psi}=\sin\frac{\theta}{2}\ket{01}+\cos\frac{\theta}{2}e^{i\phi}\ket{10}$ with probability $x$
(while the average fidelity for guessing an unknown qubit is 1/2).
Note that more generally, Alice may send states like
$\ket{\psi}=c_0\ket{00}+c_1\ket{01}+c_2\ket{10}$. However, as such coherent superpositions
between $\{\ket{00}\}$ and $\{\ket{01},\ket{10}\}$ are not present in the actual
mixed input state, this will not lead to any better transfer fidelities.
So, effectively Alice should choose to send a state
$\hat\rho = x\, \ket{\psi}_{\rm guess}\!\bra{\psi} + (1-x)\, \ket{00}\!\bra{00}$,
which most closely resembles the input mixed state.
Each trial then corresponds to a density matrix
\begin{align}
\left[\begin{array}{c|cc}
1-x & 0 & 0 \\ \hline
0 & x\sin^2\frac{\theta}{2} & x\sin\frac{\theta}{2}\cos\frac{\theta}{2}e^{-i\phi} \\
0 & x\sin\frac{\theta}{2}\cos\frac{\theta}{2}e^{i\phi} & x\cos^2\frac{\theta}{2} \\ \end{array}\right]\,.
\end{align}
By averaging over all possible $\{\theta,\phi\}$, the average output density matrix can be written as
\begin{align}
\int^{\theta=\pi}_{\theta=0}\int^{\phi=2\pi}_{\phi=0}
\left[\begin{array}{c|cc}
1-x & 0 & 0 \\ \hline
0 & x\sin^2\frac{\theta}{2} & x\sin\frac{\theta}{2}\cos\frac{\theta}{2}e^{-i\phi} \\
0 & x\sin\frac{\theta}{2}\cos\frac{\theta}{2}e^{i\phi} & x\cos^2\frac{\theta}{2} \\ \end{array}\right]
\sin\theta d\theta d\phi
=\left[\begin{array}{c|cc}
1-x & 0 & 0 \\ \hline
0 & x/2 & 0 \\
0 & 0 & x/2 \\ \end{array}\right]\,.
\label{eq:alice_1}
\end{align}
This is a reasonable result, because the fidelity between
the actual input qubit state
$\left[\begin{array}{cc} 1 & 0 \\ 0 & 0 \\\end{array}\right]$
and Alice's average qubit
$\left[\begin{array}{cc} 1/2 & 0 \\ 0 & 1/2 \\\end{array}\right]$
is 1/2.

Second, assume Alice's QND measurement gives her the qubit subspace
with probability $\eta$. In this case,
Alice will send Bob an estimated qubit state with probability $1-y$
or a vacuum state $\ket{00}\!\bra{00}$ with probability $y$.
Note that since Alice has now access to the unknown input qubit living
in the qubit subspace, she can perform an optimal qubit state estimation
(rather than just guessing a random qubit as before).
For this qubit state estimation,
Alice randomly chooses one set of orthogonal measurement bases
$\{\sin\frac{\theta}{2}\ket{01}+\cos\frac{\theta}{2}e^{i\phi}\ket{10},
\cos\frac{\theta}{2}\ket{01}-\sin\frac{\theta}{2}e^{i\phi}\ket{10}\}$.
The input qubit state $\ket{01}$ will then be projected onto
\begin{itemize}
\item $\sin\frac{\theta}{2}\ket{01}+\cos\frac{\theta}{2}e^{i\phi}\ket{10}$ with probability $\sin^2\frac{\theta}{2}$,
in which case Alice sends Bob $\sin\frac{\theta}{2}\ket{01}+\cos\frac{\theta}{2}e^{i\phi}\ket{10}$,
\item $\cos\frac{\theta}{2}\ket{01}-\sin\frac{\theta}{2}e^{i\phi}\ket{10}$ with probability $\cos^2\frac{\theta}{2}$,
in which case Alice sends Bob $\cos\frac{\theta}{2}\ket{01}-\sin\frac{\theta}{2}e^{i\phi}\ket{10}$.
\end{itemize}
Therefore, Alice's estimated qubit state is
\begin{align}
&\sin^2\frac{\theta}{2}
\left[\begin{array}{c|cc}
0 & 0 & 0 \\ \hline
0 & \sin^2\frac{\theta}{2} & \sin\frac{\theta}{2}\cos\frac{\theta}{2}e^{-i\phi} \\
0 & \sin\frac{\theta}{2}\cos\frac{\theta}{2}e^{i\phi} & \cos^2\frac{\theta}{2} \\ \end{array}\right]
+\cos^2\frac{\theta}{2}
\left[\begin{array}{c|cc}
0 & 0 & 0 \\ \hline
0 & \cos^2\frac{\theta}{2} & -\sin\frac{\theta}{2}\cos\frac{\theta}{2}e^{-i\phi} \\
0 & -\sin\frac{\theta}{2}\cos\frac{\theta}{2}e^{i\phi} & \sin^2\frac{\theta}{2} \\ \end{array}\right].
\end{align}
By averaging over all $\{\theta,\phi\}$, this time we obtain
\begin{align}
\left[\begin{array}{c|cc}
0 & 0 & 0 \\ \hline
0 & 2/3 & 0 \\
0 & 0 & 1/3 \\ \end{array}\right]\,.
\end{align}
This is also a reasonable result, because
the fidelity between the input qubit
$\left[\begin{array}{cc} 1 & 0 \\ 0 & 0 \\\end{array}\right]$
and Alice's average estimated qubit
$\left[\begin{array}{cc} 2/3 & 0 \\ 0 & 1/3 \\\end{array}\right]$
is 2/3.
Recall that the overall state sent by Alice is now
$\hat\rho = (1-y)\, \ket{\psi}_{\rm estimate}\!\bra{\psi} + y\, \ket{00}\!\bra{00}$,
again most closely resembling the input mixed state.
Thus, the final average density matrix which Alice sends to Bob is
\begin{align}
\left[\begin{array}{c|cc}
y & 0 & 0 \\ \hline
0 & (1-y)2/3 & 0 \\
0 & 0 & (1-y)/3 \\ \end{array}\right]\,.
\label{eq:alice_2}
\end{align}

Now by using Eqs.~(\ref{eq:alice_1}) and (\ref{eq:alice_2}),
the final output state, including both QND measurement results
onto the qubit and the vacuum subspaces, is
\begin{align}
\rho_{\text{out}}=(1-\eta)
\left[\begin{array}{c|cc}
1-x & 0 & 0 \\ \hline
0 & x/2 & 0 \\
0 & 0 & x/2 \\ \end{array}\right]
+\eta
\left[\begin{array}{c|cc}
y & 0 & 0 \\ \hline
0 & (1-y)2/3 & 0 \\
0 & 0 & (1-y)/3 \\ \end{array}\right].
\label{eq:rho_out}
\end{align}
Since $\rho_{\text{in}}$ in Eq.~(\ref{eq:rho_in}) and $\rho_{\text{out}}$ in Eq.~(\ref{eq:rho_out}) are diagonal, we can directly calculate the state fidelity,
\begin{align}
F_\text{state}(x,y,\eta)
&=\left[\text{Tr}\left(\sqrt{\sqrt{\rho_{\text{in}}}\rho_{\text{out}}\sqrt{\rho_{\text{in}}}}\right)\right]^2\nonumber\\
&=\left[\sqrt{\eta\left(\frac12x(1-\eta)+\frac23(1-y)\eta\right)}
+\sqrt{(1-\eta)\left((1-x)(1-\eta)+y\eta\right)}\right]^2.
\end{align}
It is maximized at $x=0$ and $y=\frac{1-\eta}{3-\eta}$,
\begin{align}
F_\text{state}^{\text{max}}(\eta)
=\left[\sqrt{\eta\cdot\frac23\left(1-\frac{1-\eta}{3-\eta}\right)\eta}
+\sqrt{(1-\eta)\left((1-\eta)+\frac{1-\eta}{3-\eta}\cdot\eta\right)}\right]^2
=1-\frac{\eta}{3}\,.
\end{align}
This result is the desired classical limit, which
correctly reproduces the extreme cases of
classical teleportation of a completely known state $\hat\rho^\perp$
(like a pure vacuum state),
$F_\text{state}^{\text{max}}=1$ for $\eta=0$, and an unknown,
pure qubit state, $F_\text{state}^{\text{max}}=2/3$ for $\eta=1$.
However, note that the above general classical bound for finite $\eta$
was no longer obtained through an average over the bounds
for the two extreme cases of classical teleportation
(as discussed at the beginning of this supplementary section,
see Eq.~(\ref{eq:threshold_1})).
Instead, it is based upon an input-output mixed-state fidelity,
for which the optimal classical procedure depends on $\eta$:
Alice should always send the known non-qubit state fraction $\hat\rho^\perp$
whenever her QND detection yields the non-qubit subspace
(i.e., whenever she obtains a total photon number in the two modes
that is less or greater than one).
So, she would actually never have to guess the qubit state
for the $1-\eta$ case, corresponding to $x=0$, even when $\eta\to 1$.
However, for the $\eta$ case, i.e., whenever Alice detects
the qubit subspace with exactly one photon in the two modes,
she would still send a known non-qubit state $\hat\rho^\perp$
with a non-zero probability $y=\frac{1-\eta}{3-\eta}$ and an estimated
qubit state only with probability $1-y$. In particular, for a very small
qubit fraction $\eta\to 0$, in up to $1/3$ of those rare $\eta$ events, Alice would
actually not estimate the qubit. Only when $\eta\to 1$,
Alice would mostly apply qubit state estimation.
Nonetheless, independent of these classical protocols
and the choice of the figure of merit
(average fidelity versus input-output mixed-state fidelity),
the classical bound is $1-\frac{\eta}{3}\equiv F_\text{thr}$
in either case.

\subsection{Input-qubit independence of the fidelity}

From the simple model below, it can be seen that
$F_\text{state}$ and $F_\text{qubit}$ (as defined in the main text) are independent of the chosen qubit state
$\ket{\psi}=\alpha\ket{01}+\beta\ket{10}$ ($|\alpha|^2+|\beta|^2=1$).
First, single-mode CV teleportation with feedforward gain $g$ and squeezing parameter $r$ transforms
a vacuum state, $\hat{\rho}_0\equiv\ket{0}\!\bra{0}$, and a single-photon state,
$\hat{\rho}_1\equiv\ket{1}\!\bra{1}$, as
\begin{align}
\hat{\rho}_{i}\equiv\ket{i}\!\bra{i}\longrightarrow
\hat{\rho}_{i}^\prime
\equiv&\int d^2\beta\hat{T}^g_q(\beta)\ket{i}\!\bra{i}\hat{T}^{g\dagger}_q(\beta) \nonumber \\
=&\sum_{n=0}^{\infty}P_i^n(g,q)\ket{n}\!\bra{n}\>\>\>(i=0,1),
\label{eq:01teleportation}
\end{align}
where $q=\tanh r$,
\begin{align}
\hat{T}^g_q(\beta)=\sqrt{\frac{1-q^2}{\pi}}\sum_{n=0}^{\infty}q^n\hat{D}(g\beta)\ket{n}\!\bra{n}\hat{D}(-\beta)
\end{align}
is a transfer operator~\cite{s02Ide}, $\hat{D}(\beta)$ is a displacement operator with an amplitude of $\beta$, and
\begin{align}
P_0^n(g,q)&=\frac{(1-q^2)(g-q)^{2n}}{(1+g^2-2gq)^{n+1}},\label{eq:0coefficient}\\
P_1^n(g,q)&=\frac{(1\!-\!q^2)(g\!-\!q)^{2n-2}}{(1\!+\!g^2\!-\!2gq)^{n+2}}
\times\left[(1\!-\!gq)^2(g\!-\!q)^2\!+\!ng^2(1\!-\!q^2)^2\right].\label{eq:1coefficient}
\end{align}
By using Eq.~(\ref{eq:01teleportation}),
teleportation of a dual-rail qubit can also be described.
The experimental input qubit state can be modeled by a mixed state of a pure qubit and the vacuum,
written as
\begin{align}
\hat{\rho}_\text{in}=\eta\ket{\psi}\!\bra{\psi}+(1-\eta)\ket{00}\!\bra{00}\quad (0\le\eta\le1).
\label{eq:model_in}
\end{align}
For an arbitrary qubit state $\ket{\psi}$,
an appropriate basis transformation defined by a unitary operator $\hat{U}$ allows to decompose
$\hat{\rho}_\text{in}$ in Eq.~(\ref{eq:model_in}) as
\begin{align}
\hat{\rho}_\text{in}^U\equiv\hat{U}\hat{\rho}_{\text{in}}\hat{U}^\dagger=\hat{\rho}_{0}\otimes\left[(1-\eta)\hat{\rho}_{0}+\eta\hat{\rho}_{1}\right].
\label{eq:decomposed_in}
\end{align}
Since the basis transformation and the teleportation process commute~\cite{s02Ide},
quantum teleportation can be discussed in this basis.
From Eq.~(\ref{eq:01teleportation}), dual-rail CV teleportation transforms $\hat{\rho}_\text{in}^U$ as
\begin{align}
\hat{\rho}_\text{in}^U\longrightarrow
\hat{\rho}_{\text{out}}^U\equiv\hat{\rho}_{0}^\prime\otimes\left[(1-\eta)\hat{\rho}_{0}^\prime+\eta\hat{\rho}_{1}^\prime\right].
\label{eq:decomposed_out}
\end{align}
Finally the output density matrix in the original basis is obtained as $\hat{\rho}_\text{out}\equiv\hat{U}^\dagger\hat{\rho}_{\text{out}}^U\hat{U}$.
Due to the properties of the fidelity~\cite{s94Jozsa},
$F_\text{state}$ and $F_\text{qubit}$ between $\hat{\rho}_{\text{in}}$ and $\hat{\rho}_{\text{out}}$ are equal to
those between $\hat{\rho}_{\text{in}}^U$ and $\hat{\rho}_{\text{out}}^U$.
Thus, from Eqs.~(\ref{eq:decomposed_in}) and (\ref{eq:decomposed_out}), we obtain
\begin{align}
F_{\text{state}}
&\!=\!P_0^0\Big\{\sqrt{(1-\eta)\left[(1-\eta)P_0^0+\eta P_1^0\right]}
+\sqrt{\eta\left[(1-\eta)P_0^1+\eta P_1^1\right]}\Big\}^2,\\
F_{\text{qubit}}
&\!=\!\frac{P_0^0\left[(1\!-\!\eta)P_0^1\!+\!\eta P_1^1\right]}
{P_0^0\!\left[(1\!-\!\eta)P_0^1\!+\!\eta P_1^1\right]\!+\!P_0^1\!\left[(1-\eta)P_0^0\!+\!\eta P_1^0\right]}.
\end{align}
These fidelities are functions of $\eta$, $q(=\tanh r)$, and $g$,
but independent of $\ket{\psi}$.
As a result, the optimal gain that maximizes the fidelity is also independent of $\ket{\psi}$.

\section{Supplementary Data}

In our experiment, quantum teleportation is performed under various conditions
to observe how each parameter affects the performance of teleportation.
First, two non-orthogonal qubits $\ket{\psi_1}\equiv(\ket{01}-i\ket{10})/\sqrt2$ and $\ket{\psi_2}\equiv(2\ket{01}-\ket{10})/\sqrt5$
are teleported at three pure squeezing parameters $r=0.71\pm0.02$, $1.01\pm0.03$, $1.56\pm0.06$ with varying classical channel gains $g=0.50$, $0.63$, $0.79$, $1.0$.
The $r$ values correspond to three different pumping levels of the OPOs (30 mW, 60 mW, and 120 mW), and
they are deduced from the measured correlation and anti-correlation of the quadrature-entangled beams.
The gain is first adjusted to unity by following the method of Ref.~\cite{s08Yukawa}, and then lowered via step attenuators in the classical channel.
Two-mode density matrices for these qubit states are obtained by dual-homodyne measurement~\cite{s12Takeda2}.
In addition to $\ket{\psi_1}$ and $\ket{\psi_2}$, the two states $\ket{01}$ and $\ket{10}$ are teleported at $r=0.71\pm0.02$ and $g=0.79$
to see if the success criterion is also satisfied for these qubit states.
Since teleportation of $\ket{01}$ and $\ket{10}$ corresponds to simultaneous teleportation of a single-mode
single-photon state and a vacuum state,
their two-mode density matrices are deduced as tensor products of two single-mode density matrices obtained by single-homodyne measurement.

\begin{figure*}[!ht]
\begin{center}
\includegraphics[width=0.75\linewidth]{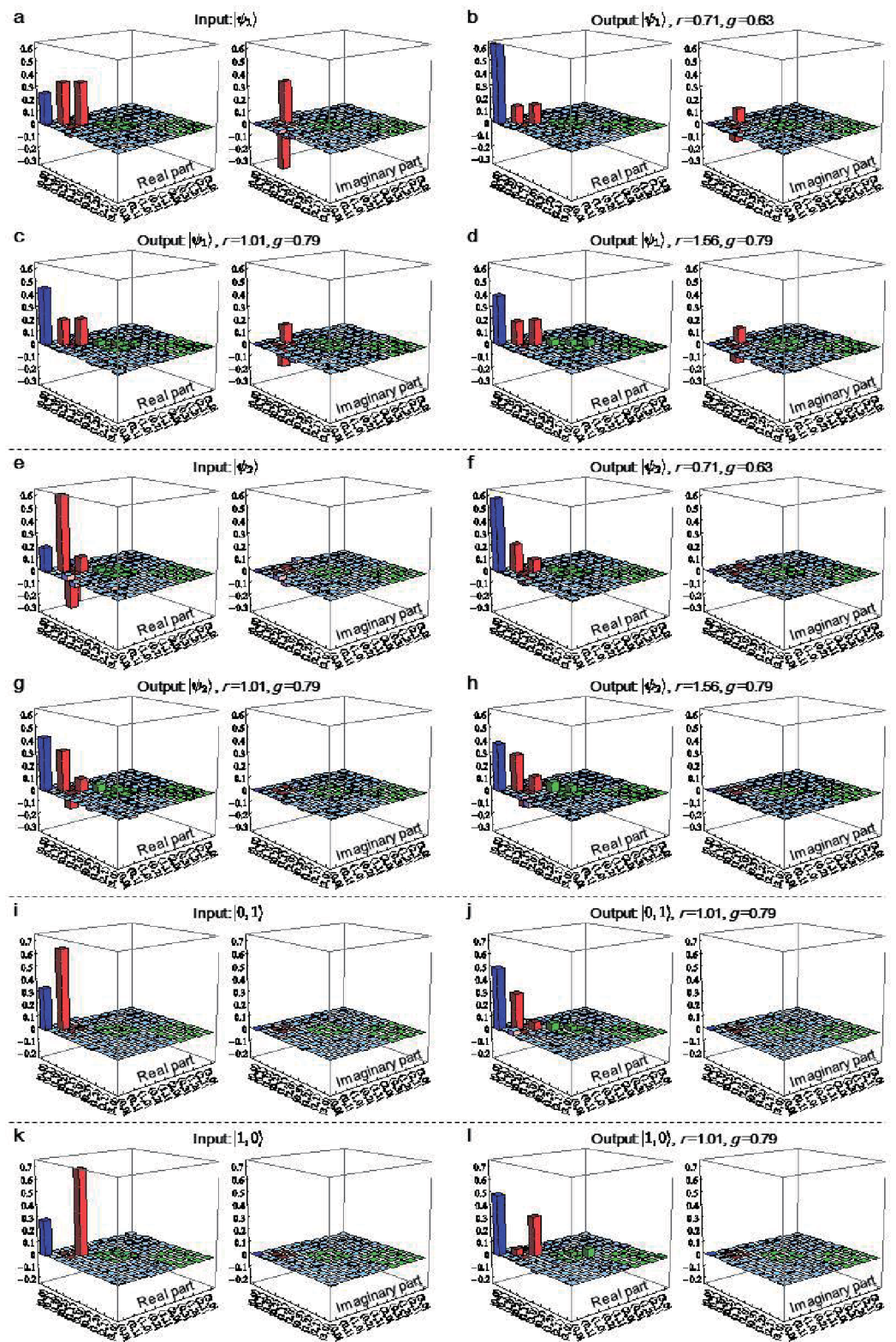}
\end{center}
\vspace{-5mm}
\caption{
{\bf Experimental density matrices.}
Blue, red and green bars correspond to vacuum, qubit and multi-photon components, respectively.
{\bf a-d.} Teleportation of $\ket{\psi_1}$ ({\bf a-c} is also included in Fig.~2 of the main text).
{\bf e-h.} Teleportation of $\ket{\psi_2}$.
{\bf i-j.} Teleportation of $\ket{01}$.
{\bf k-l.} Teleportation of $\ket{10}$.
}
\label{fig:dmatrix_sup}
\vspace{-2mm}
\end{figure*}

Experimental density matrices of gain-tuned teleportation for each input state and $r$ are summarized in Supp.~Fig.~\ref{fig:dmatrix_sup}.
For both $\ket{\psi_1}$ and $\ket{\psi_2}$, the teleported states at the higher squeezing level show more multi-photon components;
the discrepancy between the effective squeezing and anti-squeezing levels
(i.e., the deviation from a pure squeezed state) increases as $r$ increases (see,
for example, Ref.~\cite{s07Takeno}),
which adds unwanted photons to the teleportation process.
In the case of $\ket{\psi_2}$ (Supp. Figs.~\ref{fig:dmatrix_sup}{\bf e}-{\bf h}),
we can see that both the amplitude ratio of the qubit ($\ket{01}\!\bra{01}$ and $\ket{10}\!\bra{10}$)
and the phase information of the superposition ($\ket{01}\!\bra{10}$ and $\ket{10}\!\bra{01}$)
are maintained after teleportation.
Theoretically, the input and output density matrices have non-vanishing components only
in the $\{\ket{00}\}$ subspace (zero photons: vacuum),
the $\{\ket{01},\ket{10}\}$ subspace (one photon: qubit), the $\{\ket{02},\ket{11},\ket{20}\}$ subspace (total photon number two), and so on.
However, all the experimental density matrices also have small non-zero off-diagonal components which are not predicted from theory, such as $\ket{02}\!\bra{00}$.
With our simulation, we were able to confirm that this is partly due to the imperfection of the
dual-homodyne characterization of time-bin qubits~\cite{s12Takeda2}:
the quadratures measured at the two homodyne detectors are not perfectly orthogonal.
Additionally, the asymmetric teleportation process for the two orthogonal axes of the phase space might also contribute to such components
(\textit{e.g.}, asymmetric squeezing levels, gains, and homodyne visibilities for two orthogonal quadratures).

\begin{figure*}[!ht]
\begin{center}
\includegraphics[width=\linewidth]{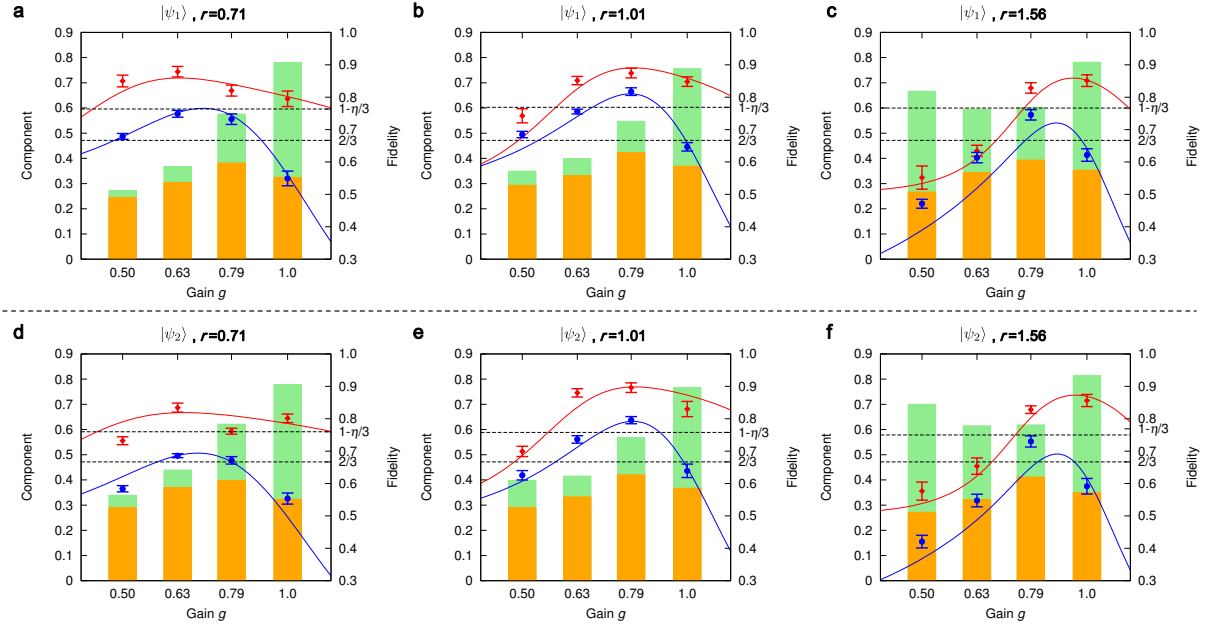}
\end{center}
\vspace{-5mm}
\caption{
{\bf Gain dependence of teleportation.}
Orange and green bars represent qubit and multi-photon components in the teleported states, respectively (the left vertical axis),
while red diamonds and blue circles correspond to $F_\text{qubit}$ and $F_\text{state}$, respectively (the right vertical axis).
Theoretical fidelity curves are also plotted with the same colors.
{\bf a-c.} Teleportation of $\ket{\psi_1}$ ({\bf b} is the same as Fig.~3 of the main text).
{\bf d-f.} Teleportation of $\ket{\psi_2}$.}
\label{fig:g_dependence_qubit2}
\vspace{-2mm}
\end{figure*}

The $g$ dependence of teleportation for $\ket{\psi_1}$ and $\ket{\psi_2}$
is plotted in Supp.~Fig.~\ref{fig:g_dependence_qubit2},
and the maximal fidelities, $F_\text{state}^\text{max}$ and $F_\text{qubit}^\text{max}$, for the best gains $g_\text{best}$ are summarized in Supp. Table~\ref{tb:r_dependence}.
No significant difference can be seen between the two input states.
The $g$ dependence clearly varies with $r$,
and the best gains $g_\text{best}$ for $F_\text{state}^\text{max}$ and $F_\text{qubit}^\text{max}$ are close to $g_\text{opt}\equiv\tanh r$.
For all $r$, high values of $F_\text{qubit}^\text{max}$ beyond the classical limit of $2/3$ are obtained.
These $F_\text{qubit}^\text{max}$ values do not show any squeezing-level dependence, and they give an average value of $\overline{F}_\text{qubit}^\text{max}=0.87\pm0.03$;
this value is comparable to that obtained in previous photonic-qubit teleportation experiments.
In contrast, $F_\text{state}^\text{max}$ takes its maximum at $r=1.01$ both for $\ket{\psi_1}$ and $\ket{\psi_2}$ (see Supp.~Table~\ref{tb:r_dependence}),
and the more rigorous Fock-space criterion $F_\text{state}>F_\text{thr}$ is satisfied only for this $r$.
So there is an optimal squeezing level for $F_\text{state}^\text{max}$,
because the increase of the multi-photon components in the teleported states
makes the overall fidelity $F_\text{state}$ deteriorate when the squeezing levels are too high.
Note that the Fock-space criterion is also fulfilled for $\ket{01}$ and $\ket{10}$ at the optimal $r=1.01$ and $g=0.79$,
as shown in Supp. Table~\ref{tb:qubit_dependence}.

In Supp.~Fig.~~\ref{fig:g_dependence_qubit2},
the $g$ dependence of $F_\text{state}$ and $F_\text{qubit}$ coincide reasonably well with the theoretical curves,
which are each calculated based on the experimental input state $\hat{\rho}_\text{in}$, the squeezing parameter $r$,
and the loss on the squeezing $l$ that minimizes the error sum of squares~\cite{s10Mista,s12Mizuta}.
The loss estimated this way ranges from $l=0.17$ to $l=0.32$.
The actual loss in our CV teleporter includes
the OPO escape efficiency $\eta_\text{OPO}=0.98$,
the propagation efficiency $\eta_\text{pr}=0.96$,
the average homodyne visibility $\eta_\text{vis}=0.99$,
and the detection efficiency $\eta_\text{det}=0.98$
given by the quantum efficiency of the photodiodes and the electronic noise.
These values correspond to $1-\eta_\text{OPO}\eta_\text{pr}\eta_\text{vis}^2\eta_\text{det}=0.11$ for the loss.
Other causes of loss and imperfection include
the unwanted offset and fluctuation of each phase locking,
the fluctuation in the beam paths due to the optical delay line,
and the high-pass filtering of the homodyne signals in the classical channel and the measurement system
(the effect of the last one can be circumvented by introducing a mode-filtering technique into the qubit generation~\cite{s12Takeda1}).
A further slight discrepancy of the experimental values from the theoretical curves
may be attributed to the fluctuation and drift of the experimental conditions (such as $r$ and homodyne visibility) during the measurement.
The larger discrepancy at $r=1.56\pm0.06$ (Supp. Figs.~\ref{fig:g_dependence_qubit2}{\bf c} and 2{\bf f})
may be explained by its relatively large error bar
(in other cases, $r=0.71\pm0.02$ and $r=1.01\pm0.03$):
the actual $r$ during the quantum teleportation process might be slightly different from the measured $r$.

\begin{table*}[!ht]
\centering
\begin{tabular}{|c|c|cc|cc|cc|cc|}
\hline
\multicolumn{2}{|c|}{} & \multicolumn{4}{c|}{Input: $\ket{\psi_1}$} & \multicolumn{4}{c|}{Input: $\ket{\psi_2}$} \\ \hline
\ \ $r$ \ \ & \ $g_\text{opt}$ \ & \ $g_\text{best}$ \ & $F_\text{state}^\text{max}$ & \ $g_\text{best}$ \ & $F_\text{qubit}^\text{max}$
& \ $g_\text{best}$ \ & $F_\text{state}^\text{max}$ & \ $g_\text{best}$ \ & $F_\text{qubit}^\text{max}$  \\ \hline
$0.71\pm0.02$ & $0.61$ & $0.63$ & $0.749\pm0.011$ & $0.63$ & $0.879\pm0.016$ & $0.63$ & $0.686\pm0.006$ & $0.63$ & $0.834\pm0.014$ \\
$1.01\pm0.03$ & $0.77$ & $0.79$ & $0.817\pm0.012$ & $0.79$ & $0.875\pm0.015$ & $0.79$ & $0.796\pm0.011$ & $0.79$ & $0.896\pm0.015$ \\
$1.56\pm0.06$ & $0.91$ & $0.79$ & $0.746\pm0.016$ & $1.0$ & $0.851\pm0.018$ & $0.79$ & $0.730\pm0.017$ & $1.0$ & $0.856\pm0.019$ \\
\hline
\end{tabular}
\caption{
{\bf Maximum fidelity at different squeezing levels.}
$g_\text{best}$ denotes the best gain among $\{0.50, 0.63, 0.79, 1.0\}$ which gives the maximum fidelity $F_\text{qubit}^\text{max}$ or $F_\text{state}^\text{max}$.
}
\label{tb:r_dependence}
\end{table*}

\begin{table*}[!ht]
\centering
\begin{tabular}{|c|c|c|}
\hline
\ Input \ & $F_\text{state}$ & $F_\text{thr}$ \\ \hline
$\ket{\psi_1}$ & \ $0.817\pm0.012$ \ & \ $0.769\pm0.004$ \ \\
$\ket{\psi_2}$ & $0.796\pm0.011$ & $0.758\pm0.005$ \\
$\ket{01}$ & $0.800\pm0.006$ & $0.785\pm0.002$ \\
$\ket{10}$ & $0.789\pm0.006$ & $0.769\pm0.002$ \\
\hline
\end{tabular}
\caption{
{\bf Fidelity for each input state.}
These values are obtained for $r=1.01$ and $g=0.79$.
The Fock-space criterion $F_\text{state}>F_\text{thr}$ is fulfilled for all the four input states.
}
\label{tb:qubit_dependence}
\end{table*}


\begin{thebibliography}{}

\bibitem{93Bennett}
Bennett, C. H. \textit{et al.} Teleporting an unknown quantum state via dual classical and Einstein-Podolsky-Rosen channels.
\textit{Phys. Rev. Lett.} {\bf 81}, 1895-1899 (1993).

\bibitem{98Briegel}
Briegel, H.-J., D\"{u}r, W., Cirac, J. I. \& Zoller, P.  Quantum repeaters: the role of imperfect local operations in quantum communication.
\textit{Phys. Rev. Lett.} {\bf 81}, 5932-5935 (1998).

\bibitem{99Gottesman}
Gottesman, D. \& Chuang, I. L. Demonstrating the viability of universal quantum computation using teleportation and single-qubit operations.
\textit{Nature} {\bf 402}, 390-393 (1999).

\bibitem{01Knill}
Knill, E., Laflamme, R., Milburn, G. J. A scheme for efficient quantum computation with linear optics.
\textit{Nature} {\bf 409}, 46-52 (2001).

\bibitem{01Raussendorf}
Raussendorf, R. \& Briegel, H. J. A One-Way Quantum Computer.
\textit{Phys. Rev. Lett.} {\bf 86}, 5188-5191 (2001).

\bibitem{97Bouwmeester}
Bouwmeester, D. \textit{et al.} Experimental quantum teleportation.
\textit{Nature} {\bf 390}, 575-579 (1997).

\bibitem{98Boschi}
Boschi, D., Branca, S., De Martini, F., Hardy, L. \& Popescu, S. Experimental realization of teleporting an unknown pure quantum state via dual classical Einstein-Podolsky-Rosen channels.
\textit{Phys. Rev. Lett.} {\bf 80}, 1121-1125 (1998).

\bibitem{01Kim}
Kim, Y.-H., Kulik, S. P., Shih, Y. Quantum teleportation of polarization state with a complete Bell state measurement.
\textit{Phys. Rev. Lett.} {\bf 86}, 1370-1373 (2001).

\bibitem{03Marcikic}
Marcikic, I., de Riedmatten, H., Tittel, W., Zbinden, H. \& Gisin, N. Long-distance teleportation of qubits at telecommunication wavelengths.
\textit{Nature} {\bf 421}, 509-513 (2003).

\bibitem{03Pan}
Pan, J.-W., Gasparoni, S., Aspelmeyer, M. Jennewein, T. \& Zeilinger, A. Experimental realization of freely propagating teleported qubits.
\textit{Nature} {\bf 421}, 721-725 (2003).

\bibitem{12Ma}
Ma, X.-S. \textit{et al.} Quantum teleportation over 143 kilometres using active feed-forward.
\textit{Nature} {\bf 489}, 269-273 (2012).

\bibitem{Luetkenhaus}
L\"{u}tkenhaus, N., Calsamiglia, J. \& Suominen, K.-A. Bell measurements for teleportation.
\textit{Phys. Rev. A} {\bf 59}, 3295 (1999).

\bibitem{12Pan}
Pan, J.-W. \textit{et al.} Multiphoton entanglement and interferometry.
\textit{Rev. Mod. Phys.} {\bf 84}, 777-838 (2012).

\bibitem{94Vaidman}
Vaidman, L. Teleportation of quantum states.
\textit{Phys. Rev. A} {\bf 49}, 1473-1476 (1994).

\bibitem{SamKimble} Braunstein, S. L. \& Kimble, H. J.
Teleportation of Continuous Quantum Variables.
\textit{Phys. Rev. Lett.} {\bf 80}, 869 (1998).

\bibitem{98Furusawa}
Furusawa, A. \textit{et. al.} Unconditional Quantum Teleportation.
\textit{Science} {\bf 282}, 706-709 (1998).

\bibitem{01Hofmann}
Hofmann, H. F., Ide, T., Kobayashi, T. \& Furusawa, A. Information losses in continuous-variable quantum teleportation.
\textit{Phys. Rev. A} {\bf 64}, 040301(R) (2001).

\bibitem{Polkinghorne}
Polkinghorne, R. E. S. \& Ralph, T. C.
Continuous Variable Entanglement Swapping.
\textit{Phys. Rev. Lett.} {\bf 83}, 2095-2099 (1999).

\bibitem{NielsenChuang}
Nielsen, M. A. \& Chuang, I. L.
Quantum Computation and Quantum Information.
Cambridge University Press (2000).

\bibitem{98Braunstein}
Braunstein, S. L. \& Kimble, H. J. A posteriori teleportation.
\textit{Nature} {\bf 394}, 840-841 (1998).

\bibitem{07Yonezawa}
Yonezawa, H., Braunstein, S. L. \& Furusawa, A. Experimental Demonstration of Quantum Teleportation of Broadband Squeezing.
\textit{Phys. Rev. Lett.} {\bf 99}, 110503 (2007).

\bibitem{01Ide}
Ide, T., Hofmann, H. F., Kobayashi T. \& Furusawa, A. Continuous-variable teleportation of single-photon states.
\textit{Phys. Rev. A} {\bf 65}, 012313 (2001).

\bibitem{11Lee}
Lee, N. \textit{et al.} Teleportation of Nonclassical Wave Packets of Light.
\textit{Science} {\bf 332}, 330-333 (2011).

\bibitem{12Takeda2}
Takeda, S. \textit{et al.} Generation and eight-port homodyne characterization of time-bin qubits for continuous-variable quantum information processing.
\textit{Phys. Rev. A} {\bf 87}, 043803 (2013).

\bibitem{03Bowen}
Bowen, W. P. \textit{et al.}, Experimental investigation of continuous-variable quantum teleportation.
\textit{Phys. Rev. A} {\bf 67}, 032302 (2003).

\bibitem{04Jia}
Jia, X. \textit{et al.}, Experimental Demonstration of Unconditional Entanglement Swapping for Continuous Variables.
\textit{Phys. Rev. Lett.} {\bf 93}, 250503 (2004).

\bibitem{10Mista}
Mi\v{s}ta Jr., L., Filip, R. \& Furusawa, A. Continuous-variable teleportation of a negative Wigner function.
\textit{Phys. Rev. A} {\bf 82}, 012322 (2010).

\bibitem{94Jozsa}
Jozsa, R. Fidelity for mixed quantum states.
\textit{J. Mod. Opt.} {\bf 41}, 2315-2323 (1994).

\bibitem{95Massar}
Massar, S. \& Popescu, S. Optimal extraction of information from finite quantum ensembles.
\textit{Phys. Rev. Lett.} {\bf 74}, 1259-1263 (1995).

\bibitem{06Zavatta}
Zavatta, A., D'Angelo, M., Parigi, V., \& Bellini, M. Remote Preparation of Arbitrary Time-Encoded Single-Photon Ebits.
\textit{Phys. Rev. Lett.} {\bf 96}, 020502 (2006).

\end{thebibliography}

\begin{thebibliography}{99}
\bibitem{s95Massar}
Massar, S. \& Popescu, S. Optimal extraction of information from finite quantum ensembles.
\textit{Phys. Rev. Lett.} {\bf 74}, 1259-1263 (1995).

\bibitem{s94Jozsa}
Jozsa, R. Fidelity for mixed quantum states.
\textit{J. Mod. Opt.} {\bf 41}, 2315-2323 (1994).

\bibitem{s02Ide}
T. Ide \textit{et al.}, Gain tuning and fidelity in continuous-variable quantum teleportation.
\textit{Phys. Rev. A} {\bf 65}, 062303 (2002).

\bibitem{s08Yukawa}
Yukawa, M., Benichi, H. \& Furusawa, A. High-fidelity continuous-variable quantum teleportation toward multistep quantum operations.
\textit{Phys. Rev. A} {\bf 77}, 022314 (2008).

\bibitem{s07Takeno}
Takeno, Y., Yukawa. M., Yonezawa H. \& Furusawa, A. Observation of $-9$ dB quadrature squeezing with improvement of phase stability in homodyne measurement.
\textit{Opt. Express} {\bf 15}, 4321-4327 (2007).

\bibitem{s12Takeda2}
Takeda, S. \textit{et al.} Generation and eight-port homodyne characterization of time-bin qubits for continuous-variable quantum information processing.
\textit{Phys. Rev. A} {\bf 87}, 043803 (2013).

\bibitem{s10Mista}
Mi\v{s}ta Jr., L., Filip, R. \& Furusawa, A. Continuous-variable teleportation of a negative Wigner function.
\textit{Phys. Rev. A} {\bf 82}, 012322 (2010).

\bibitem{s12Mizuta}
Takeda, S. \textit{et al.}
Gain tuning for continuous-variable quantum teleportation of discrete-variable states
\textit{Phys. Rev. A} {\bf 88}, 042327 (2013).

\bibitem{s12Takeda1}
Takeda, S. \textit{et al.} Quantum mode filtering of non-Gaussian states for teleportation-based
quantum information processing.
\textit{Phys. Rev. A} {\bf 85}, 053824 (2012).

\end{thebibliography}
\end{document}